# A physical zero-knowledge object-comparison system for nuclear warhead verification

Sébastien Philippe[1], Robert J. Goldston[2], Alexander Glaser[1] & Francesco d'Errico[3,4]

Zero-knowledge proofs are mathematical cryptographic methods to demonstrate the validity of a claim while providing no further information beyond the claim itself. The possibility of using such proofs to process classified and other sensitive physical data has attracted attention, especially in the field of nuclear arms control. Here we demonstrate a non-electronic fast neutron differential radiography technique using superheated emulsion detectors that can confirm that two objects are identical without revealing their geometry or composition. Such a technique could form the basis of a verification system that could confirm the authenticity of nuclear weapons without sharing any secret design information. More broadly, by demonstrating a physical zero-knowledge proof that can compare physical properties of objects, this experiment opens the door to developing other such secure proof-systems for other applications.

[1] Department of Mechanical and Aerospace Engineering, Princeton University, E-Quad, Olden Street, Princeton, New Jersey 08544, USA. [2] Princeton Plasma Physics Laboratory, Princeton University, PO Box 451, MS 41, Princeton, New Jersey 08540, USA. [3] School of Engineering, University of Pisa, Largo Lazzarino 1, 56126 Pisa, Italy. [4] School of Medicine, Yale University, 300 Cedar Street, New Haven, Connecticut 06520, USA. Correspondence and requests for materials should be addressed to S.P. (email: sp6@princeton.edu).

The next round of arms-control agreements may require a trusted verification mechanism to confirm the authenticity of items presented as nuclear warheads. Proliferation and national security concerns require, however, that such a verification mechanism reveals no information about the composition or design of these warheads[1–3]. Considerable research efforts have been directed towards the development of engineered information barriers to address this paradox[4,5]. These barriers consist of automated measurement and analysis systems that process sensitive information but only display the results of internal analysis in a binary manner (valid object and invalid object). These systems are, by their nature, at risk of electronic tampering and snooping—and their trusted implementation has so far proved difficult to realize[6–8].

Glaser, Barak and Goldston (GBG)[6] have recently proposed a different approach based on a zero-knowledge interactive proof, or protocol, for warhead verification. Physical zero-knowledge protocols can be used to perform comparisons or computations on physical data such that sensitive information is never measured and does not need to be protected afterwards[9]. They rely on the concept of interactive zero-knowledge proofs, originally developed for computational cryptographic applications[10], which have the property of yielding no knowledge beyond the validity of the assertion being proven. This property remains guaranteed only if the prover follows the protocol[11]. Attempts at cheating open the possibility of information leakage, a potential deterrent for a range of attacks[6,7]. The timely demonstration of a trusted zero-knowledge warhead inspection system would represent a breakthrough towards a verification regime targeted at deeper and multilateral cuts in nuclear arsenals. Other critical applications of physical zero-knowledge proofs could include any system where classified or personal data need to be protected, such as in forensic DNA analysis[9].

We have devised a physical zero-knowledge system, using a non-electronic neutron differential radiography technique, to perform reproducible object comparison without ever acquiring data about the objects being compared. We employ this system to demonstrate experimentally the validity of key aspects of the protocol for warhead verification proposed by GBG. In a real inspection, this protocol requires the existence of at least one reference warhead, for example, retrieved from active delivery systems[12,13] at the inspector's designation. Using simple radiographic test objects, we find experimentally that items practically identical to the reference item can be confirmed to be genuine while inspectors gain no knowledge about their geometry and composition. Furthermore, items differing significantly from the reference item can be discriminated.

## Results

**A zero-knowledge object-comparison system.** Our technique compares the neutron radiographic profile of an inspected item against that of a reference item at various energies (in particular 14 MeV) to confirm that the items are identical within experimental accuracy (Fig. 1). In the case of nuclear weapons, such data are highly sensitive. We circumvent this problem by recording the radiographs of the inspected items on sets of previously exposed superheated emulsion (bubble) detectors[14] preloaded with the 'complement' of the transmission image of the reference item (or any item claimed to be similar to the reference item; Fig. 1a). In this case, the 'complement' implemented by the host, the owner of the items, is defined as the preload that will result in the total fluence recorded in each detector being the same as if no item had been present when the inspected item is the same as the reference item—we call this total count $N_{max}$. Such a radiograph with a flat profile of $N_{max}$ and Poisson noise conveys zero information. We emphasize that the inspector is not present while the host prepares the preloads. In the GBG protocol the inspector is offered multiple sets of detectors preloaded by the host and then randomly selects which to use either with the reference item(s) or with the inspected item(s). This prevents the host from matching with certainty a modified preload to a spoof warhead. The use of non-electronic detectors in the protocol also avoids the risks of electronic tampering by the host once the preload has been selected and of electronic snooping by the inspector to attempt to learn the preload pattern.

**The inspection protocol.** We envision that the GBG protocol could be realized the following way: first, the host declares $(n-1)$ items as 'treaty accountable items' that are to be inspected and compared with a reference item (for a total of $n$ items). Here we assume a single reference item, but there could be more than one in principle. The host claims that all $n$ items are identical. Then, the parties must agree to the appropriate level of statistics for the inspection. This includes agreeing on $N_{max}$ as well as any statistical test to be applied to the data. Probabilities of passing a test depend on random choices made by the inspector and the measurement noise, which are independent between each invocation of the inspection test. Finally, we assume that parties have agreed on procedures for chain of custody, perimeter monitoring and other security measures related to the conduct of the inspections.

*Initialization*

1. The host produces a collection of detectors. He measures and records their neutron efficiency curves. The inspector then chooses a subset of these detectors to be used in the inspections, and can take home the others and independently confirm their calibration curves.

2. The host takes a large number of radiographs of non-reference items at $m$ different angles, and $l$ different incoming neutron energies minimizing any systematic errors in the preloads. The radiographs can be taken on non-reference items to avoid difficulties associated with maintaining chain of custody for the reference item, which is the only item inspectors believed to be genuine before inspections start.

*Preparation of preloads.* The host, without the inspector present, prepares $n$ preloaded detector sets $p_{i,j}^k$ with $k = 1...n$, each corresponding to the complement image of the treaty accountable items (claimed to be identical to the reference) at a particular angle $\theta_i$, for $i = 1...m$, and particular energy $E_j$, for $j = 1...l$.

*Inspection*

1. The inspector calls for an orientation $\theta_i$ of the items and energy $E_j$ of the neutron beam.

2. The host presents the corresponding $n$ preloaded detector sets $p_{i,j}^1, \ldots, p_{i,j}^n$ to the inspector (including the associated calibration data). The inspector chooses a random assignment of the detector sets to the reference item and $(n-1)$ inspected items.

3. The items are exposed to the neutron source and their radiographs are recorded on the preloaded detectors. Both parties monitor the source fluence—each with their own apparatus. This ensures the inspector that a measurement is taking place, and that the agreed fluence corresponding to $N_{max}$ has been provided.

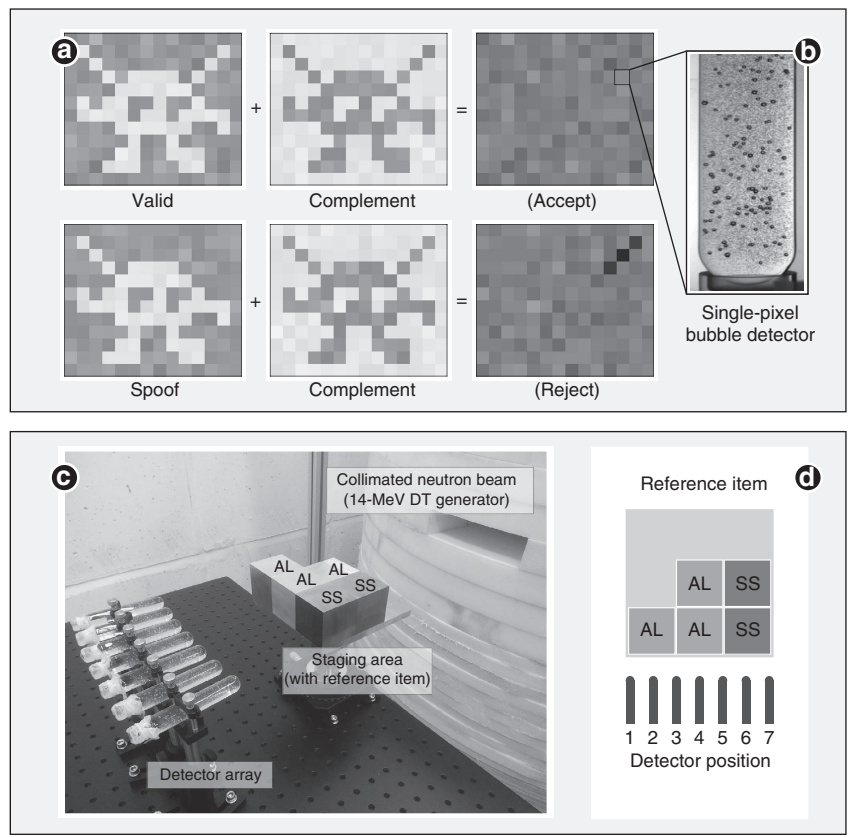

**Figure 1 | Experimental realization of a physical zero-knowledge object comparison system.** (**a**) Concept of zero-knowledge differential neutron radiography using superheated droplet detectors. Items are exposed to a neutron beam and their 2D transmission radiographs are recorded on detectors preloaded with the complement radiograph (including Poisson noise) of a reference item. If the item is valid (identical to the reference), the final radiograph is identical to the expected exposure if no object had been present (in the ideal implementation the root mean square deviation from the expected bubble count, $N_{max}$, is solely because of Poisson noise, $(N_{max})^{0.5}$). If the item is a spoof with an experimentally significantly different radiograph, some characteristic features appear in the final radiograph (the results are no longer zero-knowledge) and the inspector rejects the proof. Each pixel represents the bubble count from a single superheated droplet (bubble) detector. (**b**) Picture of an irradiated superheated droplet (bubble) detector. Some metastable droplets vapourized and expanded into macroscopic bubbles after a neutron interaction. The bubble count reflects the total fluence delivered to the detector. (**c**) One-dimensional experimental realization with superheated droplet detector array, reference item (black box hood not shown) on staging area and aperture of a 14 MeV collimated neutron beam. The set-up is placed in a room shielded with borated concrete walls. A fast neutron counter (not shown) monitored the source fluence along the axial direction. (**d**) Composition and pattern of the reference item with detector positions (AL, aluminium; SS, stainless steel).

4. After the inspection of each item, the number of macroscopic bubbles in each detector is counted. Both parties compute the result of the agreed upon test for the hypothesis that the distribution of the counts is $n$ independent Poisson random variables each with expectation $N_{max}$.
5. Some detectors, as selected by the inspector, are recompressed (to remove the bubbles, preparing them for re-use) and re-exposed to the source to verify their functionality and specified calibration.

*Test repetition.* The inspection phase can be repeated to increase arbitrarily the soundness of the proof. The inspector can decide to re-test items at the same angle and energy (for example, if a particular item has failed the test), or continue to test the items at another $(\theta_i, E_j)$ pair. By this means, the protocol can meet arbitrarily stringent overall false-positive and -negative rates by accumulating statistics. Strategies to minimize the effort to achieve specific overall type I and type II errors will be the subject of a future publication.

In this realization, we assume that the inspector follows the protocol described here. All actions of the inspector, unlike those of the host, are performed in public view of both parties. In particular, the inspector has only access to the bubble count in each detector using an agreed-upon bubble-counting technique. She does not, for example, have any possibility to measure the neutron-induced activation level of the bubble detectors.

**Properties of the protocol.** Zero knowledge: At the end of the protocol, if the host behaved honestly, we expect the total bubble count in each detector to be Poisson-distributed with expectation value $N_{max}$, the sum of a matching preload and radiograph— equal to the total count if no item had been present between the source and the detectors. For the zero-knowledge property to be conserved, neither the signal nor the noise may carry information.

For a particle beam experiment, where many particles would strike a detector for every recorded interaction, and in the case where discrete counts are independent, random and occurring at a constant rate, the counts should be distributed according to a Poisson distribution[15–17]. Recalling that the sum of two independent Poisson random variables is a Poisson random variable[18], we obtain that the sum of the radiograph and its matching complement is Poisson-distributed with parameter $N_{max}$—that is, both the mean and variance of this distribution are $N_{max}$, in the limit of high accuracy of calibration and preload as discussed infra.

Therefore, as long as the items are identical, the verifier observes $n$ independent draws from a Poisson($N_{max}$) random variable. Since the inspector could have sampled this *a priori*, she does not learn anything.

Soundness and completeness: The soundness and completeness of the protocol are based on the assumption that two items with significantly different geometry and/or composition result in a difference in the expected bubble counts. 'Significant differences' are defined by the parties within the treaty protocol.

Neutron counts in the detectors are affected by the geometry and physical properties of the item. For example, neutron opacity depends on the total neutron cross-section, which is energy- and nuclide-dependent; neutron intensities and their distributions at large angles depend on elastic and inelastic scattering, fission and (n,2n) reaction cross-sections, again energy- and nuclide-dependent; geometry and fissile material composition affect the neutron multiplication factor, which is also energy-dependent.

Exposing the inspected item to different neutron energies allows verifying that both geometry and material composition are not significantly different. Energies at which neutrons are highly penetrating (such as 14-MeV neutrons) or where large differences in fission cross-sections[19], or large resonances in the total cross-section[20] exist, are of particular interest. It is possible to probe the object at an energy lower than the detector energy threshold for bubble nucleation to emphasize differences in the fission signature[21].

Formally, we assume that if an item X is not significantly different from the reference item, then it passes inspections at all angle–energy ($\theta_i$, $E_j$) pairs with probability of at least $(1 − \alpha)$, where $\alpha$ is the false-positive rate. Furthermore, we make the physical assumption that there exists $\beta$ such that for any reference item X and any object Y created by the adversary, if Y significantly differs from X, then there exists an orientation–energy pair ($\theta_i$, $E_j$) such that testing with these parameters the inspector will declare the item to pass the inspection with probability of at most $\beta$.

If the host is cheating, at least $s \geq 1$ spoofs are presented for inspection. Suppose the host also provide $p_s$ spoof preloads (defined as the complement of the spoof item) and $(n − p_s)$ valid preloads (complement of the reference item). One of the following cases must apply:

In the first case, $s \neq p_s$, then either at least one spoof is tested against a valid preload or at least one valid item is tested against a spoof preload. In either case, the cheating host evades detection with probability of at most $\beta$.

In the second case, $s = p_s$, one of the items will be tested against the wrong type of preload with probability $1 − \frac{1}{C(n,s)}$ where $C(n,s)$ is the number of $s$ combinations from the set of $n$ items. Thus, the adversary evades detection with probability of at most $1/C(n,s) + \beta(1 − 1/C(n,s))$. In this case, the probability of evasion is always greater than $\beta$. Therefore, such a cheating strategy should always be preferred over $s \neq p_s$.

Since $s \leq n − 1$, the probability of evasion is maximized when $s = p_s = 1$ or $s = p_s = n − 1$. Since $s$ is under host control, we assume that the host will present $n − 1$ spoofs and their associated spoof preloads. In this case, the probability to evade detection is bounded by $1/n + \beta(1 − 1/n)$. Therefore, for the protocol to be sound $1/n + \beta(1 − 1/n) < 1 − \alpha$, which can also be written as $\beta < 1 − \alpha n/(n − 1)$. In the worst case, $n = 2$ (only one reference and one inspected item), $\beta < 1 − 2\alpha$, which also means that $\alpha < 1/2$.

As noted above, the probability of evading detection can be arbitrarily reduced by re-testing. This would also detect cases where the host provides $n$ preloads with counts between those of the spoof and the reference item. The closer $\beta$ is to $1 − \alpha n/(n − 1)$, the more repetition will be needed to ensure a high soundness level, however. Efficient and acceptable testing strategies based on the values of $\alpha$, $\beta$ and the soundness and completeness levels negotiated between parties will need to be developed. For $n = 10$ and $\beta = 0.1$, the probability of evasion would be equal to 0.19, which may be sufficient to deter the host from using the strategies we described.

**Experimental realization.** For this proof of principle, and to provide a physical basis to the GBG protocol, we inspected simple items in one orientation with one neutron beam energy fixed at 14 MeV. We represented the reference warhead and the inspected items by combinations of 2-inch cubes of aluminium and steel. The cubes attenuate differently 14 MeV neutrons (44% and 27% for 2.54 cm of steel and aluminium, respectively) and can be arranged in different patterns. Figure 1c shows the experimental set-up, as well as the cube pattern used to represent the internal components of the reference item. The item also included a 'black box' aluminium hood to mimic the fact that inspected objects would likely be in containers, hiding their appearance. In a real implementation, to prevent disclosure of the preload, the detectors would also be covered with visually opaque material until the irradiation of the items was complete. In addition, a fast neutron counter monitors the source fluence to assure that it is the agreed value to produce the total number of bubbles expected in the detectors in the absence of a test item.

To prove the validity and usefulness of our technique, we tested both its ability to convey zero knowledge in cases where the item presented for inspection is identical to the reference item (valid case) and its robustness to detect cases in which the cube pattern had been altered, referred to as spoofs.

For an interactive zero-knowledge proof applied to nuclear warhead verification, the burden of proof[22] falls on the host (the owner of the inspected items) to demonstrate that his claim is robust while his secrets are protected. As mentioned earlier, to prepare the inspection, the host must calibrate the measurement system and appropriately characterize the reference item. The calibration will be checked by inspectors through test irradiations in the absence of interposed items. In our demonstration, after characterizing the fluence response function, $f$, of the detector system, we acquired 10 radiographs of the reference item and summed the results at each detector position. Each exposure lasted ∼4 min; without an interposed item, the corresponding fluence would produce a total value of $N_{max}$ ∼ 1,200 bubbles.

Once the reference item is fully characterized, the host can produce the necessary complementary radiographs, by irradiating the bare detectors to a fluence equal to $f^{−1}(N_{max}) − f^{−1}(N_{ref,i})$ where $f^{−1}$ is the inverse of the detector response function (counted bubbles as a function of counts from the reference fast neutron detector), and $N_{ref,i}$ the bubble count obtained from the radiograph of the reference item at every position $i = \{1,...,7\}$. $f$ is a nonlinear function of the number of bubbles present in a bubble detector. The nonlinearity is because of bubbles occulting one another in the photographic images used for bubble counting (see Fig. 1b). Individual exposure was kept at $N$ ∼ 120 to avoid large nonlinear bubble-counting response. The analysis reported hereafter is based on the number of bubbles present in the detectors, which is somewhat larger than the number counted, obtained by correcting the data for bubble occultation (See Methods). In practice, when preparing detectors for individual exposure, we used a linear approximation to $f$, which introduced a <2% error in the preloads.

It is necessary to correct for the slight difference in gain between bubble detectors and for the effect of the location of any given detector. The number of bubbles is not exactly $N_{max}$ in each detector exposed to the same fluence when no item is present. Using data from 10 different exposures of the detectors at a fluence $f^{-1}(N_{max})$, we found that our detectors differed in efficiency with 3.7% $1-\sigma$ dispersion. We corrected our results to take these variations into account. This calibration itself introduces a variance of $N_{max}$ in the final experimental results. The total expected variance for our measurements consists of (1) this calibration variance, $N_{max}$, (2) the variance of the radiograph of the reference item (or valid item), $N_{ref}$, (3) the variance of the preload, $N_{max} - N_{ref}$ and (4) the variance of the subsequent exposure of the inspected item, $N_{item}$. The calibration variance (1) and the reference radiograph variance (2) can be driven down as low as desired by the inspector and the host, respectively. The former will improve the sensitivity of the measurement, while the latter will prevent this variance from being a source of information leakage when a large number of valid items is inspected. In the limit where only variance sources (3) and (4) play a role, and an valid item is presented, therefore, $N_{item} = N_{ref}$, then the expected distribution is Poisson with the mean value $N_{max}$ and variance $N_{max}$.

**Experimental results.** We performed measurements for five different scenarios, one with an inspected item identical to the reference item and four scenarios with spoofs whose cube patterns differed from that of the reference item. During the inspections, the patterns were placed under the 'black box' hood and their radiographs recorded on a set of seven bubble detectors preloaded with the complement radiograph of the reference item. Each scenario was repeated 10 times and the bubble counts were summed to achieve an effective $N_{max} = 1,191$. The use of 10 measurements allowed us to use the variance of the mean to determine the error bars in the measurement experimentally. In general, this variance was found to be indistinguishable from the measurement as expected from Poisson statistics (see Methods). Figure 2a summarizes all our experimental results (the results uncorrected for occultation are shown in Methods). Figure 2b shows good agreement between Monte–Carlo neutron transport simulations of the experiment that calculated the neutron fluence in the detectors using the code Monte Carlo N-particle[23], and the experimental results. These data also support the validity of our nonlinear occultation model.

In the 'valid' case, data are randomly distributed around $N_{max}$ and the measured mean variance from $N_{max}$, $\sum_{i=1}^{7} \frac{(N-N_{max})^2}{7} = 2646$, is in good agreement with the predicted value of $\sum_{i=1}^{7} (\frac{(2N_{max}+N_{ref})^{0.5}}{7})^2 = 2964$. This is consistent with an inspector gaining no knowledge about the transmission profile of the reference item from this measurement. In the first diversion scenario (Spoof1), the cubes making up the inspected item are swapped from left to right compared with the reference item. This change can be easily identified in the data. The next two diversion scenarios (Spoof2 and Spoof3) consisted of altering the reference item pattern by either adding (increased attenuation) or removing (decreased attenuation) a single steel cube. Again, both cases are clearly identified as different from the reference item with very high confidence. Finally, in the last case (Spoof4), an aluminium cube replaced a steel cube on the right side (small reduction of the attenuation). This case could be considered an acceptable spoof detection, with $P = 0.013$, depending on the agreement between host and inspector. The data, however, suggest that a tighter criterion or repeated testing be used for this magnitude of $P$.

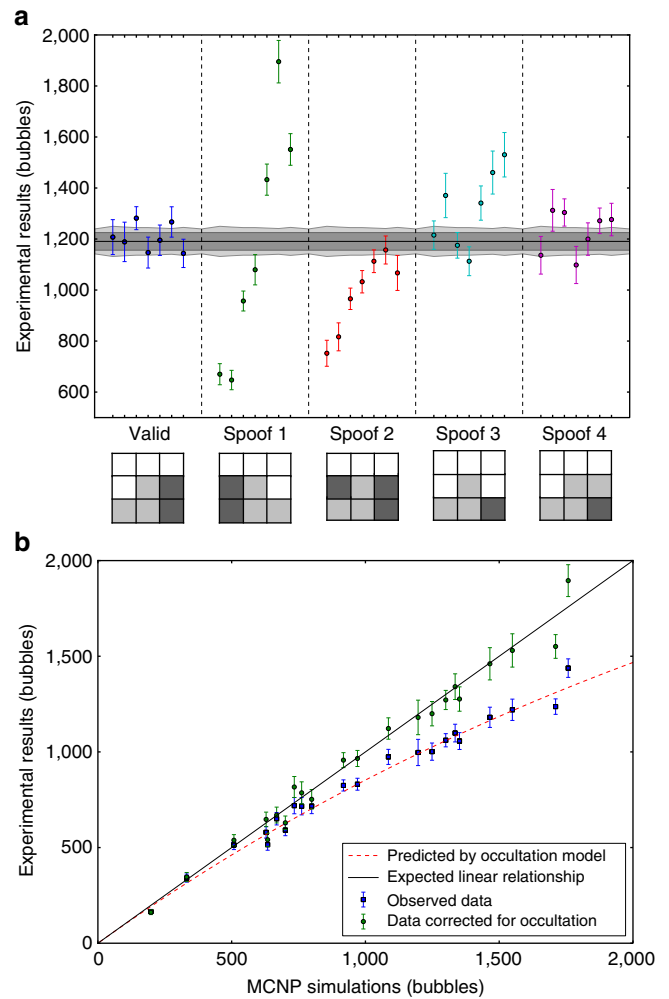

**Figure 2 | Experimental evidence of a practical physical zero-knowledge proof.** (**a**) Experimental results for the five inspection scenarios investigated (valid item and four spoofs denominated Spoof1 to Spoof4). The item patterns are represented for each scenario with white as empty, light grey as aluminium and dark grey as steel. The results are corrected for the nonlinear counting effect. Error bars represent one s.e.m. calculated at each detector position from the 10 measurements performed for each scenario and the calibration data (obtained from 10 measurements with no item). The light grey band around $N_{max}$ represents the expected error from a valid item $(2N_{max}+N_{ref})^{0.5}$. The dark grey band corresponds to the minimum achievable value with $N_{max}$ bubbles, $(N_{max})^{0.5}$. The test statistic $T = \sum_{i=1}^{7} \frac{(N-N_{max})^2}{2N_{max}+N_{ref}}$ was compared with a $\chi^2$ distribution with 7 degrees of freedom, giving the following results for each scenario: Valid, $T = 6.60$, $P = 0.474$; Spoof 1, $T = 453.35$, $P < 10^{-16}$; Spoof 2, $T = 132.72$, $P < 10^{-16}$; Spoof 3, $T = 87.80$, $P = 3.10^{-16}$; Spoof 4, $T = 17.88$, $P = 0.013$. (**b**) Observed and deduced true bubble count, from all scenarios at every position with interposed objects, versus corresponding Monte Carlo simulations obtained from the computational model of the experiment. The red dashed curve is obtained from our bubble occultation model using the calibration data with no interposed objects (see Methods). Error bars represent one s.e.m. calculated from measurements and calibration data.

## Discussion

This experiment demonstrates the key aspects of the viability of non-electronic differential fast neutron radiography to perform zero-knowledge object comparison, establishing a physical basis for the nuclear warhead verification approach proposed by GBG. It constitutes the demonstration of a physical zero-knowledge proof protocol for a practical high-security application that can compare physical properties of objects. We have shown that the combination of pre-loading and direct measurement on an honest object gives the desired null result, with a well-understood level of noise. We have identified areas where the noise can be reduced to improve the technique. We have also explored the capability of this technique at the level of $N_{max} = 1{,}200$ bubbles for detecting diversions. Further research, at higher bubble counts, with different neutron energies, and with special nuclear materials, will be needed to assess the capabilities of the system for the full range of relevant diversion scenarios.

## Methods

**Experimental set-up.** The radiographs of the test items are performed using 14 MeV neutrons from a D–T neutron generator (Thermofisher B320) with a yield of $\sim 10^8$ neutrons per second. The generator is placed inside a borated polyethylene cylinder (1 m high × 1.2 m in diameter) with a 5 cm high and 17 degree-wide fan-shaped collimating slot. A single row of seven specially made superheated droplet neutron (or 'bubble') detectors are placed $\sim 110$ cm away from the source, behind the test item. The items are mounted on a stand affixed to a honeycomb aluminium optical table allowing precise and reproducible positioning (staging area in Fig. 1). A fast neutron counter (Eljen EJ-410 zinc sulfide scintillator) monitors the generator output during irradiation by detecting neutrons emerging from a channel at the bottom of the polyethylene cylinder. The complete apparatus is placed in a room shielded by borated concrete walls at the Princeton Plasma Physics Laboratory (Plainsboro, USA).

The detectors developed for this application[14,19] consist of glass tubes filled with emulsions of superheated octafluorocyclobutane ($C_4F_8$, fluorocarbon C-318) 100 μm diameter droplets homogeneously dispersed in a viscous aqueous gel matrix (4,000 per $cm^3$). At 21 °C the detectors are only sensitive to neutrons above 1 MeV and insensitive to gamma rays. When a metastable droplet vapourizes, because of a sufficiently energetic neutron interaction, it expands into a stable bubble about six times larger in diameter. The detectors can be irradiated several times in series, recording the total fluence to which they were exposed (pre-loadable property). If an isostatic pressure of 500 p.s.i. is applied to the aqueous gel for $\sim 10$ min, the bubbles are recondensed into superheated droplets.

All bubble detectors used in this study were manufactured from the same batch. Fourteen detectors were chosen out of twenty-one after being irradiated several times at every position for $\sim 6$ min to compare their relative behaviour. The chosen detectors exhibited a 3.7% $1-\sigma$ dispersion and were calibrated to allow determination of their absolute efficiency (of the order of $4 \times 10^{-4}$ bubbles per crossing neutron in an active volume of 5.3 $cm^3$).

Data were acquired in two ways to study the effect of recompression on the detector response. One detector batch corresponding to 5,000, 10,000, up to 25,000, fast neutron counts were irradiated, measured, recompressed and then irradiated to the next higher level. A second batch was irradiated, counted and then further irradiated to provide data at 2,500, 7,500, and so on, counts. The recompression was shown to have no detectable effect on the measurements.

Because the energy threshold of bubble nucleation is temperature-dependent, we monitored the temperature of the detectors throughout the study. The experimental hall was temperature-controlled and the detector temperature exhibited minor variations over all measurements (20.79 ± 0.34 °C).

After irradiation, photographs of the detectors were taken using a commercial instrument (BDR-III reader, Bubble Technology Industries (BTI)) at multiple angles. The bubbles were then counted with the associated BTI software with settings modified to accommodate our detector tubes. The results were cross-checked with visual counting using large printed copies of the images. The main challenge of these techniques stems from the correct segmentation of bubbles that overlap in the two-dimensional (2D) image.

**Bubble occultation model.** Before starting any measurements, we carefully characterized the fluence response of our detector system, $b_{vis} = f(n)$, where $b_{vis}$ is the number of bubbles recorded by the reading system and $n$ the counts from the fast neutron detector. A nonlinear behaviour emerged, because of the bubble-counting method used in the experiments. Bubbles are counted from a 2D $(y,z)$ projection of the volume perpendicular to the $x$ axis (direction of light) and parallel to the vertical $z$ axis of the cylindrical detector. Thus, bubbles at the same $z$ and $y$ positions may occult one another. (Note that these data do not necessarily imply that BTI bubble detectors read with BTI equipment produce the level of non-linearity we observed. These results apply only to our own detectors with tailored settings to accommodate our specific material and geometric conditions.) To account for this nonlinear effect, we used a simple particle interaction model. For an observer placed far away from the detector and looking in the $x$ direction, a new bubble will be visible only if its image does not cross an old bubble. Thus, we can assume that for each new bubble, on average the observer only sees $e^{-\frac{b_r \sigma d}{V}}$ bubbles, where $V$ is the counting volume, $d$ is a characteristic width of the cylinder along the line of sight, $\sigma$ the effective occultation cross-section and $b_r$ the real number of bubbles at the time a new bubble is created. Averaging over the cylinder indicates that $d$ should be taken to be $4/\pi$ times the cylinder radius. We expect the number of real bubbles to be a linear function of the neutron count in the fast neutron detector because only a very small fraction of the droplets have vapourized ($\sim 0.6\%$ for 4 min irradiation), and the fractional volume occupied by the vapourized droplets is very small ($\sim 0.25\%$ volume displacement). Thus, we assume that $\frac{db_r}{dn} = \varepsilon$. This gives us the rate at which visible bubbles accumulate as a function of neutron irradiation, $\frac{db_{vis}}{dn} = \varepsilon e^{-\frac{b_r \sigma d}{V}}$, and integrating this expression, using $b_r = \varepsilon n$, we obtain $b_{vis} = \frac{V}{\sigma d}(1 - e^{-\frac{\varepsilon \sigma n d}{V}})$. This model has efficiency $\varepsilon$ at low fluence and saturates, when $b_r$ goes to infinity, at $b_{vis} = \frac{V}{\sigma d}$.

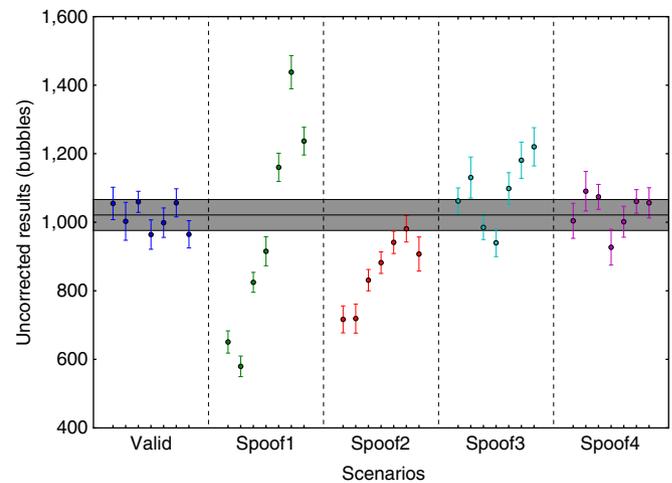

**Figure 3 | Uncorrected experimental results.** Uncorrected experimental results for the five inspection scenarios investigated and defined in the Results section (valid item and four diversions denominated Spoof1 to Spoof4). Data points are presented with one s.e.m. error bars calculated at each detector position from the 10 measurements performed for each scenario and the calibration data (obtained from 10 measurements with no item). The grey band is $\pm (2N_{max})^{0.5}$ centred on $N_{max}$.

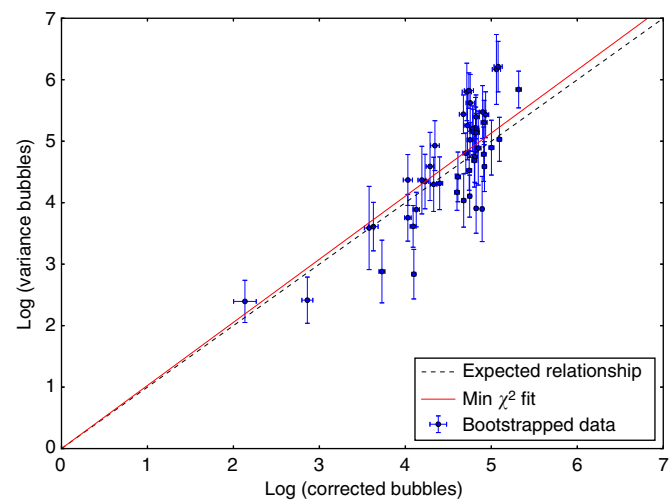

**Figure 4 | Analysis of the superheated droplet detector variance.** Plot of the logarithm of the variance versus the logarithm of the average number of true (corrected) bubbles. The data and one s.d. error bars were obtained by bootstrapping the corrected experimental results (1,000 samples per data point).

Calibration data, obtained at increasing exposures without interposed items, were fitted to this model to obtain the true number of bubbles $b_r$ from $b_{vis}$. (In practice, this calibration procedure also compensates for the small effects because of droplet vapourization and occupied volume.) Data obtained later from the detector preloads confirmed the good fit and reproducibility of this effect. We find that the effective cross-section $\sigma$ is equivalent to a disk with a radius of the order of twice the radius of a single bubble ($\sim 300\,\mu m$), effectively corresponding to two bubbles touching one another.

**Uncorrected inspection results.** In the main section, the inspection results were corrected for the nonlinear effect introduced by the reading technique. Figure 3 presents the uncorrected data. These results were directly obtained from the readout, and calibrated to account for spatial and detector gain differences as explained in the main text.

**Detector variance.** The zero-knowledge property of our protocol depends on its ability to carry information neither in the signal nor in the noise of the measurements. Since we are measuring radiation, the sum of the detector counts (preload and radiograph of inspected item) should follow Poisson statistics. Thus, at each position we expect the data obtained for all detectors to have a variance equal to their bubble count. To verify that the noise of our detector varies as $\sqrt{N}$, we generated synthetic data from our experimental results using the bootstrapping method[24]. The bootstrapping method consists of estimating population parameters by randomly sampling with replacement from an approximating distribution (here the empirical population obtained from our experimental data). We used 10 measurements for each inspection scenario (there are 5) as well as 10 radiographs of the reference item and 10 preloads (complement radiographs) as starting data sets. A measurement corresponds to 7 data points for each position. We generated 1,000 synthetic measurements for each case to obtain the bootstrapped mean and variance with their respective error bars. The results presented in Fig. 4 shows good agreement with the expected $\sqrt{N}$ noise from Poisson statistics. Note that here we are using $b_r$ deduced from the nonlinear model. The good agreement indicates that the occultation process and its correction do not introduce significant additional noise at the level of nonlinearity explored.

**Data availability.** Source data for figures are available from the corresponding author upon request.

## References

1. Comley, C. *et al. Confidence, Security & Verification: the Challenge of Global Nuclear Weapons Arms Control* (Atomic Weapons Establishment, 2000).
2. National Academy of Sciences CISAC, *Monitoring Nuclear Weapons and Nuclear-Explosive Materials: an Assessment of Methods and Capabilities* (National Academies Press, 2005).
3. Stone, R. Not-seeing is believing. *Science* **344,** 1436–1437 (2014).
4. Spears, D. (ed.) *Technology R&D for Arms Control* (US Department of Energy, Office of Nonproliferation Research and Engineering, 2001).
5. Evans, N. Software Development and Authentication for Arms Control Information Barriers. In *FM 2015: formal Methods* 581–584 (Springer International Publishing, 2015).
6. Glaser, A., Barak, B. & Goldston, R. A zero-knowledge protocol for nuclear warhead verification. *Nature* **510,** 497–502 (2014).
7. Philippe, S., Barak, B. & Glaser, A. Designing protocols for nuclear warhead verification. In *Proc. 56th Annual INMM Meeting* (Indian Wells, CA, USA, 2015).
8. Fisch, B., Freund, D. & Naor, M. Secure Physical Computation using Disposable Circuits. In *Theory of Cryptography* 182–198 (Springer, 2015).
9. Fisch, B., Freund, D. & Naor, M. Physical Zero-Knowledge Proofs of Physical Properties. In *Advances in Cryptology—CRYPTO 2014* 313–336 (Springer, 2014).
10. Goldwasser, S., Micali, S. & Rackoff, C. The knowledge complexity of interactive proof-systems. *SIAM J. Comput.* **18,** 186–208 (1989).
11. Goldreich, O. *Foundations of Cryptography: Basic Tools* (Cambridge University Press, 2000).
12. Marleau, P. *et al. Report on a zero-knowledge protocol tabletop exercise* (SAND2015-5075, Sandia National Laboratory, 2015).
13. Yan, J. & Glaser, A. Nuclear warhead verification: a review of attribute and template systems. *Sci. Global Security* **23,** 157–170 (2015).
14. d'Errico, F. Radiation dosimetry and spectrometry with superheated emulsions. *Nucl. Instrum. Methods Phys. Res. B* **184,** 229–254 (2001).
15. Knoll, G. F. *Radiation Detection and Measurement* (John Wiley & Sons, 2010).
16. Lyons, L. *Statistics for Nuclear and Particle Physicists* (Cambridge University Press, 1986).
17. James, F. *Statistical Methods in Experimental Physics* 2nd edn (World Scientific, 2006).
18. Lehmann, E. L. & Romano, J. P. *Testing Statistical Hypotheses, 3rd edn* (Springer, New York, 2005).
19. Goldston, R. J. *et al.* Zero-knowledge warhead verification: system requirements and detector technology. In *Proc. 55th Annual INMM Meeting* (Atlanta, GA, USA, 2014).
20. Chen, G & Lanza, R. C. Fast neutron resonance radiography for elemental imaging: theory and applications. *IEEE Trans. Nucl. Sci.* **49,** 1919–1924 (2002).
21. Yan, J. & Glaser, A. Two-color neutron detection for zero-knowledge nuclear warhead verification. In *Proc. 56th Annual INMM Meeting* (Indian Wells, CA, USA, 2015).
22. Kaplow, L. Burden of proof. *Yale L.J* **121,** 738–859 (2012).
23. A general Monte Carlo N-particle (MCNP) transport code, version 6.1.0. Los Alamos National Laboratory, mcnp.lanl.gov.
24. Efron, B. & Tibshirani, R. J. *An Introduction to the Bootstrap* (CRC press, 1994).

## Acknowledgements

Princeton University, Yale University and the Princeton Plasma Physics Laboratory received support for this research through DOE/NNSA's Consortium for Verification Technology, DE-NA 0002534. Financial support was also provided by the John D. and Catherine T. MacArthur Foundation and the Carnegie Corporation of New York. We thank B. Barak for his advice on the protocol construction and assumptions, A. Carpe and the Health Physics team of the Princeton Plasma Physics Laboratory for their aid during the conduct of the experiments, M. Gattas-Sethi (Yale University) for manufacturing the detectors, Z. Mian, M. Kütt and three reviewers for providing helpful comments on the manuscript. All simulations were run on Princeton University's High Performance Cluster.

## Author contributions

S.P. designed and built the apparatus, simulated and conducted the laboratory experiments. R.J.G. and A.G. mentored S.P. and contributed, along with S.P., to the elaboration of the protocol and the analysis of the data. F.d'E. developed the detectors and participated in their characterization. All contributed to the manuscript.

## Additional information

**Competing financial interests:** The authors declare no competing financial interests.

**Reprints and permission** information is available online at http://npg.nature.com/reprintsandpermissions/

**How to cite this article:** Philippe, S. *et al.* A physical zero-knowledge object-comparison system for nuclear warhead verification. *Nat. Commun.* 7:12890 doi: 10.1038/ncomms12890 (2016).

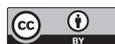